\documentclass[nofootinbib,superscriptaddress,twocolumn,aps,prc,showpacs,10pt]{revtex4-1}
\usepackage{verbatim}
\usepackage{amsmath}
\usepackage{amssymb}
\usepackage{graphicx}
\usepackage{mathrsfs}
\usepackage{color}

\begin{document}

\title{Counting statistics in finite Fermi systems: illustrations with the atomic nucleus}
\author{Denis Lacroix} \email{lacroix@ipno.in2p3.fr}
\affiliation{Institut de Physique Nucl\'eaire, IN2P3-CNRS, Universit\'e Paris-Sud, Universit\'e Paris-Saclay, F-91406 Orsay Cedex, France}
\author{Sakir Ayik}
\affiliation{Physics Department, Tennessee Technological University, Cookeville, TN 38505, USA}

\date{\today}

\begin{abstract}
We analyze here in details the probability to find a given number of particles in a finite volume inside a normal or superfluid finite system. 
This probability, also known as counting statistics, is obtained using projection operator techniques directly linked to the characteristic 
function of the probability distribution. The method is illustrated in atomic nuclei. The nature of the particle number fluctuations from small 
to large volumes compared to the system size are carefully analyzed in three cases: normal systems, superfluid systems and superfluid systems 
with total particle number restoration. The transition from Poissonian distribution in the small volume limit to Gaussian fluctuations as the number of 
particles participating to the fluctuations increases, is analyzed both in the interior and at the surface of the system. While the restoration 
of total number of particles is not necessary for small volume, we show that it affects the counting statistics as soon as more than very few particles 
are involved. 
\end{abstract}


\maketitle

\section{Introduction}

The fluctuations of the particle number inside a finite volume of a system gives insight in the nature of the particles and on their mutual interaction. 
It is also a key concept to understand static and transport properties in Fermi or Bose systems or to study the transition 
from the microscopic to the macroscopic regime \cite{Lan80}. In recent years, important efforts were made to better understand these fluctuations
and especially the second moments of the fluctuations  in atomic gases \cite{Ast07,Cas08} especially to study the BEC-BCS cross-over. 
These theoretical studies were motivated by the experimental progresses of Ref. \cite{Gre05} that were followed by a series of observations 
giving access directly to the probability to have $N$ particles in a finite volume \cite{Tor10,San10,Wes10}, 
the so-called counting statistics. In this case, it was possible to measure  local density fluctuations and more specifically the 
suppression/enhancement compared to an ideal Fermi gas for Fermions/Bosons.  This has led 
for instance to the introduction of the concept of super-Poissonian (Bosons) and sub-Poissonian (Fermions) probability distributions. 

When the number of particles contained in the system is large, as it is often the case in atomic physics, 
many properties of the fluctuations can be understood by considering the infinite number of particles limit. 
This is for instance the case in the different work \cite{Ast07,Cas08,Kla11,Bel07}. 
In the case of mesoscopic systems, where the number of constituents is not large 
enough to justify the infinite particle number limit, finite-size effects can play a significant role.  This is for 
instance the case in atomic nuclei where the number of fermions (protons and neutrons) 
can vary from very few to several hundreds. 

The aim of the present work, is to calculate not only the second moments of the particle number fluctuations in a finite volume of a system, 
but the full counting statistics. We will then study its properties when the system becomes superfluid and/or when the volume is located in the interior of the 
system or at its surface. Finally we will underline the effect of restoring the $U(1)$ symmetry associated to the conservation of the total particle number 
when the system is in a superfluid phase. Surprisingly enough and as far as we know, such counting statistics has never been directly obtained 
in atomic nuclei. 

To extract the probability distribution in a finite volume, we will use the generating function technique \cite{Bel07}. The generating function 
is intimately connected to the projection operator  approach \cite{Rin80,Bla86} that is standardly used nowadays in the nuclear many-body context.  
 The problem  of selecting a finite volume turns out to be rather similar to the problem 
addressed recently in nuclear reactions where the projection operator technique 
has been used \cite{Sim10,Sca13,Sek14,Sek16,Ver19,Bul19}. 
We will use here this technique as a starting point focusing on the static properties 
of a single Fermi system.  Since the projection operator approach is already well documented, we 
concentrate here on its link with the generating function. The technique is illustrated in nuclear systems 
described by the nuclear density functional approach eventually accounting for superfluid effects \cite{Ben03}.

\section{Counting statistic: general background}

We consider a system of $A$ fermions described by a wave-function $| \Phi \rangle$. For the moment
we do not specify the type of trial wave-function we use. Our goal is to obtain the probability distribution of
the number of particle $N$ in a finite volume $\Omega$. For this we use the projection operator technique. Since this 
technique has become a standard practical tool in recent years, we only recall here some important aspects.  

To study local properties, it is convenient to introduce local fields operators $\{ \Psi^\dagger_\alpha({\bf r}) ,  \Psi^\dagger_\alpha({\bf r}) \}$ where $\alpha$ denotes the different quantum numbers associated to the single-particle states, eventually $\alpha = (\sigma, \tau, \cdots )$  where $\sigma$ and $\tau$ are the spin and isospin components. The operator 
that counts the number of particles in a volume $\Omega$ is given by:
\begin{eqnarray}
\hat N_\Omega  &=& \sum_\alpha \int_\Omega   \Psi^\dagger_\alpha({\bf r})  \Psi_\alpha({\bf r}) d{\bf r} =  \sum_\alpha \int  \Theta_\Omega( {\bf r} ) \Psi^\dagger_\alpha({\bf r})  \Psi_\alpha({\bf r}) d{\bf r} . \nonumber
\end{eqnarray}  
In the last integral, the sum is extended to the whole volume and we introduced the function $\Theta_\Omega({\bf r})$ that is one 
if ${\bf r}$ belongs to the volume $\Omega$ and zero otherwise. We can then introduce the set of operators $\hat P_\Omega(N)$ that project on the particle number $N$:
\begin{eqnarray}
\hat P_\Omega (N) &=& \int \frac{d\varphi} {2\pi} e^{-i(N- \hat N_\Omega) \varphi} . \label{eq:projom}
\end{eqnarray}
The probability distribution is deduced from: 
\begin{eqnarray}
P_\Omega (N) &=& \langle \Phi |\hat P_\Omega (N) |\Phi \rangle=\int \frac{d\varphi} {2\pi} e^{-i N \varphi} \langle \Phi | \Phi (\varphi) \rangle. \label{eq:four}
\end{eqnarray}  
where we implicitly assumed that we have $\langle \Phi | \Phi \rangle=1$ and where we have introduced the notation 
$| \Phi(\varphi) \rangle \equiv e^{+i \hat N_\Omega \varphi } | \Phi \rangle$.  For the sake of compactness,  when no confusion is possible,  we will in the following 
omit $\Omega$ and for instance simply write $\hat N_\Omega=\hat N$ and $P_\Omega (N) = P(N)$.  

The connection between the projection technique and standard probability theory can be made by realizing 
that the function $F(\varphi) =  \langle \Phi | \Phi (\varphi) \rangle$ is the generating function of the moments of $\hat N_\Omega$. We indeed 
have:
\begin{eqnarray}
F(\varphi) = \langle e^{i\varphi \hat N_\Omega} \rangle = 1 + i \varphi \langle \hat N_\Omega \rangle- \frac{1}{2} \varphi^2  \langle \hat N^2 _\Omega\rangle + \cdots \label{eq:mom}
\end{eqnarray} 
We see from Eq. (\ref{eq:four}) that the probability distribution is nothing but the Fourier transform of $F(\varphi)$. Inversely, $F(\varphi)$ can be obtained from the inverse Fourier 
transform $F(\varphi) = \sum_N e^{i\varphi N} P(N)$.   

In practice, especially when trying to find approximate forms of the probability, it is also convenient to define the cumulant characteristic function, denoted by $G(\varphi)$. 
Following Ref. \cite{Bel07}, we introduce it as $e^{-G(\varphi)} = F(\varphi)$. The function $G(\varphi)$ is the generating function of the cumulants and we have now:
\begin{eqnarray}
G(\varphi) = 1 + i  \varphi \langle N \rangle - \mu_2 \varphi^2/2 + \cdots   \label{eq:cum}
\end{eqnarray}  
where $\mu_2 =\langle N^2 \rangle - \langle N \rangle^2$. 
We recall in table \ref{tab:gener}, the forms of the generating functions for selected probability distributions that will be useful in this work.   
\begin{table}[htbp]
   \centering\begin{tabular}{|l|c|c|}
  \hline 
  & $F(\varphi) $   & $G(\varphi)$ \\
  \hline
&&\\
Binomial & $(1+p(e^{i\varphi}-1))^N$ &$ -N \ln(1+p(e^{i\varphi}-1))$  \\
&&\\
\hline 
&& \\
Poisson & $e^{\lambda(e^{i\varphi} - 1)}$& $-\lambda (e^{i\varphi} - 1)$\\ 
&& \\
\hline
&& \\
Gaussian & $e^{i\varphi \lambda -\frac{1}{2} \mu_2^2 \varphi^2}$ & $-i\varphi \lambda  + \frac{1}{2} \mu_2^2 \varphi^2$ \\
&& \\
  \hline
\end{tabular}
\caption{Generating function for the moments and cumulants for Binomial, Poisson and Gaussian distribution using the convention introduced in the text.  
We have used the notation $\lambda=\langle N\rangle$ and, for the binomial distribution, $p$ is defined through 
$\lambda=p N$. For the Poisson distribution, we have $\lambda=\mu_2$ and for the binomial distribution $\mu_2 = N p (1-p)$.}
\label{tab:gener}
\end{table}

\subsection{Generating functions for quasi-particle vacuum}

For the moment, we did not specified the properties of the many-body state $| \Phi \rangle$. In the following, we will assume  that this state is a quasi-particle 
vacuum associated to the set of quasi-particle creation operators $\alpha_k$.  We use the standard conventions of Ref. \cite{Rin80} and write these quasi-particle operators 
in a single-particle basis associated to the set of creation/annihilation operators as $(a^\dagger_i, a_i)$ using the $(U,V)$ matrix as:
 \begin{eqnarray}
\alpha^\dagger_k &=& \sum_l \left[ U_{lk} a^\dagger_l + V_{lk} a_{l} \right].
\end{eqnarray} 
We assume further that the state takes the simplified form:
\begin{eqnarray}
| \Phi \rangle &=& \prod_{n>0} \left[u_n +v_n a^\dagger_n   a^\dagger_{\bar n}  \right] | - \rangle , \label{eq:bcslike}
\end{eqnarray}
that could be obtained for any quasi-particle vacuum using the Bloch-Messiah-Zumino decomposition \cite{Blo62,Zum62}.  Here $ | - \rangle $ is the particle vacuum. The pair of creation operators $(a^\dagger_n, a_n)$ are associated  to the wave-functions $\{ \phi_n({\bf r}) , \phi_{\bar n}({\bf r})   \}$. Expression (\ref{eq:bcslike}) implies that the $U$ and $V$ matrices 
simplify such that the only non-zero components are $U_{nn} = U_{\bar n \bar n} = u_n$ and $V_{n \bar n} = v_n, ~~V_{\bar n n} = -v_n$. 

Under the application of the projector, we first see that the state $| \Phi(\varphi) \rangle$ is given by:
\begin{eqnarray}
| \Phi(\varphi) \rangle &=& \prod_{n>0} \left[u_n +v_n b^\dagger_n (\varphi)  b^\dagger_{\bar n} (\varphi)  \right] | - \rangle , \nonumber
\end{eqnarray}
where
\begin{eqnarray}
b^\dagger_i (\varphi) &=& e^{i\hat N_\Omega \varphi} a^\dagger_i e^{-i\hat N_\Omega \varphi}= \sum_{j} R_{ij}(\varphi) a^\dagger_j .\nonumber
\end{eqnarray}
$R$ defines the transformation between the original basis and the rotated basis in gauge-space.  In the specific case considered here, we have \cite{Sca13,Ver19,Bul19}:
\begin{eqnarray}
R_{ij}(\varphi) &=& \delta_{ij} + O_{ij} (e^{i\varphi} - 1),
\end{eqnarray} 
where the $O_{ij}$ coefficients are the components of the overlap matrix $O$  between two single-particle states 
in the volume $\Omega$,  $O_{ij} = \int_\Omega \phi^*_i({\bf r}) \phi_j({\bf r}) d^{3}{\bf r}$. 
In practice, once the matrix $R$ is known, the characteristic function $F(\varphi)$ can be computed using the technique proposed in Ref. \cite{Rob09,Ber12} based on the Pfaffians. 
We then have (assuming that the number of state $n>0$ with $v_n \neq 0$ is $L$):
\begin{eqnarray}
F(\varphi) &=&  \frac{(-1)^L}{\prod^L_{n} v^2_n} {\rm pf}\left[ 
\begin{array}{cc}
A    &  B(\varphi)   \\
-B^T (\varphi)     &  A^\dagger  
\end{array}
\right]  \equiv  {\rm pf} \left[ {\cal M} \right] \label{eq:pfaf}
 \end{eqnarray}  
 with $A=V^T U$ and $B=V^T R^T(\varphi) V^*$.  In the following, all numerical applications have been obtained using the package of Ref. \cite{Gon11}.  
 
 \subsubsection{The case of diagonal overlap matrix}
 \label{sec:diag}
 
To get better insight in the formula given above, it is interesting to consider specific situations.  Let us for instance assume that the overlap matrix is diagonal
 with $O_{nn} = O_{\bar n \bar n} = p_n$. This situation would happen for instance if one considers a spherical system with 
 a set of particles in a single $j$ shell and if each particle is associated 
 to a given angular momentum projection $m$.  Then, the matrix $R$ becomes also diagonal with:
 \begin{eqnarray}
R_{nn} (\varphi) = R_{\bar n \bar n}(\varphi)  = r_{n}(\varphi) = 1+p_n (e^{i\varphi} - 1). \nonumber
\end{eqnarray} 
The matrix ${\cal M}$ becomes $4\times4$ bloc diagonal, where, in a given block associated to the pair $(n,\bar n)$, we have:
\begin{eqnarray}
{\cal M}_n &=&\frac{(-1)}{v^2_n}  \left[ 
\begin{array}{cccc}
0   &  -u_n v_n  & r_n(\varphi) v^2_n & 0\\
v_n u_n      &  0 & 0 &  r_n(\varphi) v^2_n \\
 - r_n(\varphi) v^2_n & 0 & 0 & v_n u_n \\
 0 &  - r_n(\varphi) v^2_n & -u_n v_n & 0 
\end{array}
\right] . \nonumber
\end{eqnarray}
Using the fact that ${\rm pf}\left[ {\cal M} \right]  = \prod_{n>0}  {\rm pf}\left[ {\cal M}_n \right]$, we obtain: 
\begin{eqnarray}
F(\varphi) &=&  \prod_{n>0} \{ u^2_n  + v^2_n[1+ p_n (e^{i\varphi} - 1)]^2\} , \nonumber \\
&\equiv &\prod_{n>0} F_n(\varphi) . \nonumber
\end{eqnarray}

There are a number of remarks that can be made:
\begin{itemize}
  \item The fact that the characteristic function becomes  a product of characteristic function  in each $(n,\bar n)$ sector  reveals  the fact that the probabilities 
to have $N$ particles in $\Omega$  can be constructed from the convolution of the independent probabilities $P_n(N_n)$ to have either $N_n=0$, $1$ or $2$ particles in
the volume $\Omega$, where the particles are taken from the pair $(n,\bar n)$. Starting from the expression of $F_n$, we immediately see that the probabilities to take 
$0$, $1$ or $2$ particles from the pair $n$ equal:
\begin{eqnarray}
\left\{
\begin{array}{l}
P_n(0) = u^2_n  + (1-p_n)^2 v^2_n,  \\
\\
P_n(1) = 2(1-p_n) p_n v^2_n ,   \\
\\
P_n(2) =  p^2_n v^2_n.   
\end{array}
\right.
\label{eq:p012}
\end{eqnarray}
  \item The cumulant generating function for diagonal overlap is then given by $G(\varphi)= -\sum_k \ln(F_n(\varphi))$. 
When the volume becomes infinitesimally small, meaning here small compared to the inter-particle distance $d$, all $p_n$ become very small such that 
\begin{eqnarray}
G(\varphi) &\simeq& - \sum_{n>0} \ln \left[ 1 + 2 v^2_n p_n (e^{i\varphi} - 1) \right] \nonumber \\
&\simeq& -  \langle N_\Omega \rangle (e^{i\varphi} - 1) , \nonumber 
\end{eqnarray}
where we have recognized the average number of particles in the volume given by $ \langle N_\Omega \rangle = 2 \sum_{n>0} v^2_n p_n$. We therefore recover
that the probability distribution for very small volume  becomes a Poisson distribution (see table \ref{tab:gener} and discussion in Ref. \cite{Lan80}).
\item In the limit of infinite volume, i.e. a volume much larger than the system size, then the overlap matrix is automatically diagonal with $p_n=1$ for all $n$ and we recover the standard formula (see for instance \cite{Lac09,Ben09,Dug09}):
\begin{eqnarray}
F(\varphi) &=&  \prod_{n>0} [ u^2_n  + v^2_n e^{2i\varphi} ]. \nonumber
\end{eqnarray}
Starting from this expression and assuming that all $v^2_n \ll 1$, we can again recover that the distribution becomes a Poisson distribution.  

\item In this simple diagonal approximation,  we see that the probability to have a given number of particles in the volume $\Omega$ 
will results in picking up independently particles from the different pairs $n$. Let us denote by $N^{{(e)}}_\Omega$ the number of particles 
in a given event ${(e)}$. This quantity decomposes as:
\begin{eqnarray}
N^{{(e)}}_\Omega &=& \sum_n N^{{(e)}}_n. \label{eq:nlambda}
\end{eqnarray}  
Each $N^{(e)}_n$ can be obtained by sampling the independent probabilities given by Eq. (\ref{eq:p012}). The technique of direct statistical sampling 
discussed here provides actually a straightforward statistical approach to get the counting statistics when the overlap matrix is diagonal. This approach 
is an alternative in this case to the use of the generating function\footnote{It is worth mentioning that the approach can be generalized for non-diagonal 
overlap matrix. After diagonalization of the overlap matrix, it was shown that the characteristic function can also be written as a product (see Eq.  (53-55) of Ref. \cite{Bul19}) 
of generating and therefore can be simulated using a set of independent variables.}. We note that another statistical method was proposed in Ref. \cite{Ver19}. 
A second important remark is that Eq. (\ref{eq:nlambda}) here is written as a sum of 
statistically independent variables $\{ N^{{(e)}}_n \}$ and, as a consequence, due to the central limit theorem, whatever are the probability distributions 
of the individual $N^{{(e)}}_n$ , as soon as a sufficient number of pairs will contribute, the probability distribution of  $N^{{(e)}}_\Omega$  will tend to 
a Gaussian distribution. As we will see below, this is the case even if the overlaps are not diagonal and/or the symmetry associated to the total particle number 
is restored. 

\end{itemize}

\begin{figure}[htbp]
\includegraphics[width=0.9\linewidth]{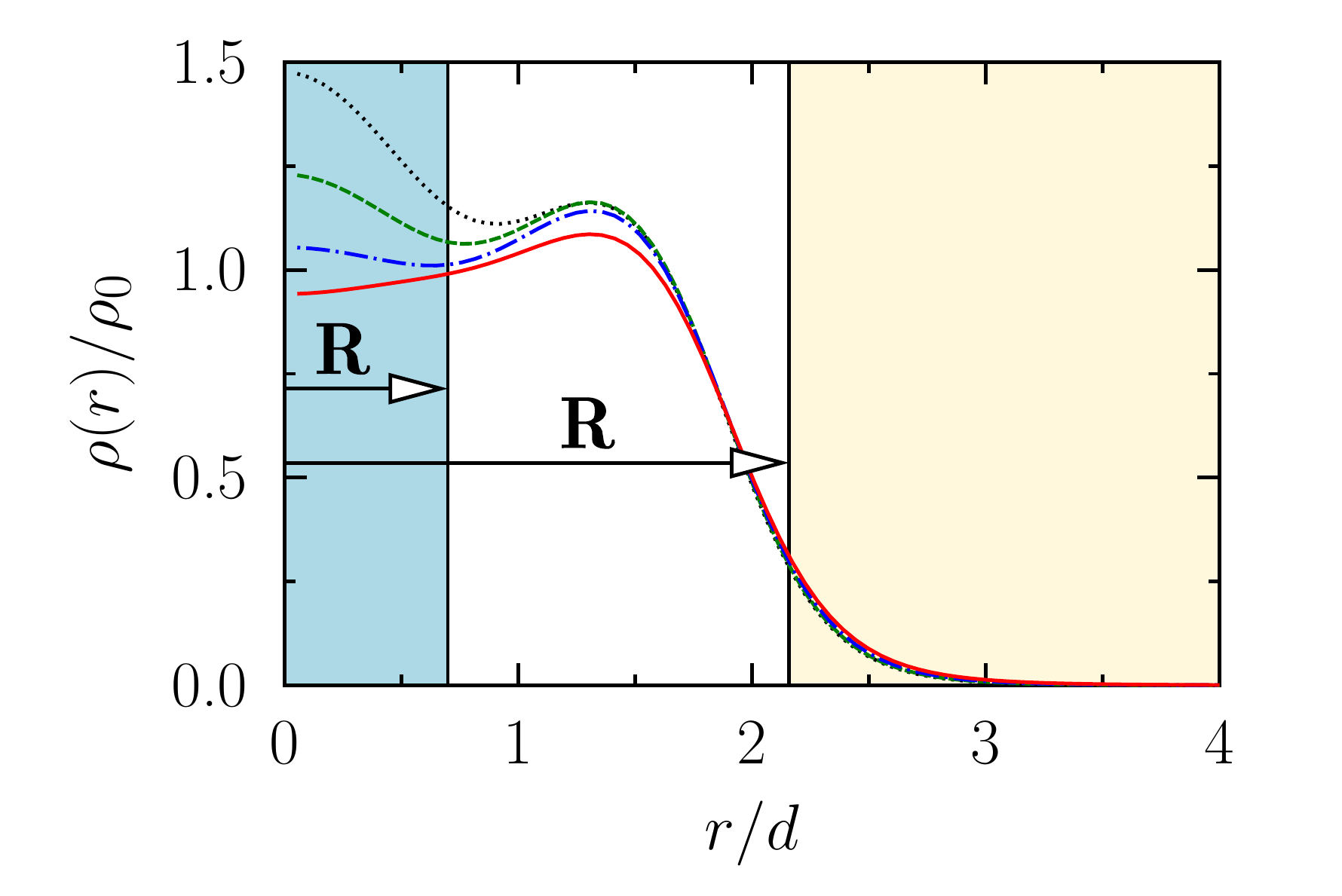}
\caption{Illustration of the local one-body density $\rho(r)$ 
as a function of $r/d$ in a $^{40}$Ca nucleus  obtained for the case of $\Delta/\Delta \varepsilon = 0.01$ [considered here as the no pairing limit] (black dotted line),
$0.5$  (green short dashed line),  $1.0$ (blue dot-dashed line) and $2.0$ (red solid line). The two colored areas illustrate the two types of volumes 
that will be considered in this work. The blue shaded area on the left indicates a volume $\Omega$ that corresponds to a sphere centered at $r/d=0$ with 
radius $R$ ($|r| < R$), while the yellow shaded area on the right indicates the volume $\Omega$ outside of the sphere ($|r|>R$). Note that two values of $R$ 
are shown for display purpose of the two shaded areas. } 
\label{fig:densA40} 
\end{figure}  
\begin{figure*}[htbp]
\includegraphics[width=\linewidth]{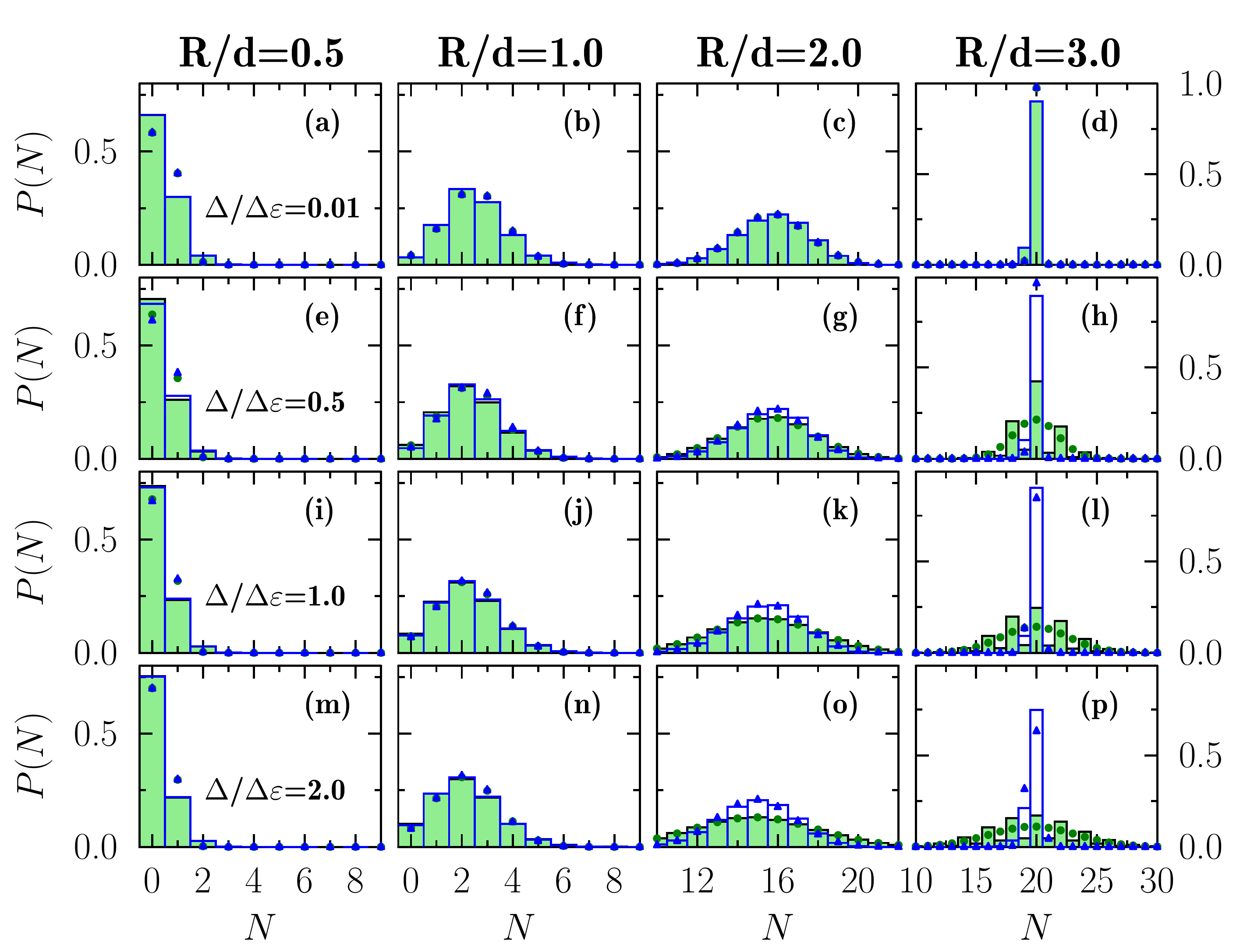}
\caption{Evolution of the probability distribution $P(N)$ of one of the species (proton or neutron) in a $A=40$ system
for different pairing gaps and different radius $R$ of the finite volume (defined as $|r|< R$). 
The pairing used in panels (a-d), (e-h), (i-l), (m-p) are respectively equal 
to $\Delta/\Delta \varepsilon = 0.01$ (no pairing limit), $\Delta/\Delta \varepsilon = 0.5$, $\Delta/\Delta \varepsilon = 1.0$ and 
$\Delta/\Delta \varepsilon = 2.0$. Each column corresponds to a given value of $R/d$ indicated on the top. 
In each panel, the green shaded area and the blue solid line
correspond to the probability without and with total particle number projection respectively. 
The green circles  and blue triangles represent the corresponding 
truncated discretized Gaussian approximation (TDGA).  } 
\label{fig:proball} 
\end{figure*}

\subsection{Probabilities with total particle number restoration}

While very useful to treat superfluid systems, the use of quasi-particles leads in general
to a breaking of the $U(1)$ symmetry 
and the number of particles in the total volume is only fixed in average. However, in many situations, the system under interest 
has a fixed number of particles and the $U(1)$ symmetry-breaking will add a spurious contribution to the counting statistics. 
A similar situation is encountered  in nuclear reactions when considering collisions involving at least one superfluid 
system \cite{Sca13,Reg18,Reg19}. A way to access to the counting statistics while 
 getting rid of the spurious contribution is to perform simultaneous projections on both 
the total particle number $A$ and on the number of particles in the volume $\Omega$.  The projector on the total 
particle number corresponds to Eq. (\ref{eq:projom}) when the volume $\Omega$ is taken as the total volume $\Omega \rightarrow + \infty$
and, for the sake of clarity, we will denote by $\hat A$ the operator associated to the total number of particles and $\theta$ the associated gauge angle.  

The discussion made previously can be easily generalized. For instance, one can write the probability, noted $Q(N,A)$ to have $N$ particles in the volume $\Omega$ 
together with $A$ particles in the complete space as:
\begin{eqnarray}
Q(N,A) &=& \int \frac{d\varphi} {2\pi}   e^{-i N \varphi } \int \frac{d\theta} {2\pi} e^{ -i \theta A} F(\varphi,\theta) \label{eq:twoproj}
\end{eqnarray}   
where the generating function is now given by:
\begin{eqnarray}
F(\varphi,\theta) &=& \langle e^{i\varphi \hat N_\Omega} e^{i\theta  \hat A} \rangle \equiv \langle \Phi | \Phi(\varphi, \theta) \rangle.   \label{eq:gen2}
\end{eqnarray}
This generating function generates now the different moments $\langle \hat N^k_{\Omega} \hat A^l \rangle$ for all integer values of $(l,k)$.
To compare with the previous case where the number of particle is not restored, we will introduce the normalized probability $P(N)$ defined 
as\footnote{Note that this probability would also correspond to the probability obtained using Eq. (\ref{eq:four}) assuming that the 
state $| \Phi \rangle$ identifies with the normalized projected HFB states:
\begin{eqnarray}
| PBCS \rangle &\equiv& \frac{1}{\sqrt{\langle \Phi | \hat P_\infty (A) |\Phi \rangle}} \hat P_\infty (A)  |\Phi \rangle
\end{eqnarray} 
where $| \Phi\rangle $ is given by Eq. (\ref{eq:bcslike}) and $ \hat P_\infty (A)$ denotes the projector on the particle number $A$ 
given by Eq. (\ref{eq:projom}) when $\Omega$ identifies with the full space.}:
\begin{eqnarray}
P(N) &=& \frac{Q(N,A)}{\sum_N Q(N,A)} \label{eq:probnorm}
\end{eqnarray} 
such that $\sum_N P(N) = 1$. Except the extra numerical effort to perform two integrations on gauge angle, the calculation is not more 
complicated than in the single-projection case. For instance, the characteristic function can also be calculated using expression (\ref{eq:pfaf})
with the difference that the $R$ matrix depends on both $\varphi$ and $\theta$ with components given by: 
\begin{eqnarray}
R_{ij}(\varphi, \theta) = \left[ \delta_{ij} + O_{ij} (e^{i\varphi} - 1)\right]e^{i\theta}.
\end{eqnarray}
Although, in the general case, the numerical integration should be performed to extract the counting statistics in a finite volume including restoration of the total particle 
number,  as a follow up to section (\ref{sec:diag}), we illustrate in appendix \ref{app:double} some simple situations where the probabilities can be worked out 
analytically.  

\section{Applications}

We now give examples of direct estimates of the counting statistics in a finite Fermi system for different volume size $\Omega$ 
with and without the restoration of the total number of particles. The atomic nucleus is an interesting test bench 
since the number of particles can vary from very few to several hundreds and finite-size effects can play a non negligible 
role.

\subsection{Initialization of superfluid self-bound nuclei}

To illustrate the method, we consider a self-bound  nucleus described within the nuclear DFT framework. To simplify the discussion, we assume 
no Coulomb and spin-orbit effects and a simplified Skyrme interaction with only $t_0$ and $t_3$ terms. The following parameters values 
are used: $t_0=-1916.1 ~{\rm MeV.fm}^{3}$, $ t_3 = 13368.6 ~{\rm MeV. fm}^{3(\alpha+1)}$ and $\alpha = 0.3024$. These parameters lead for infinite 
symmetric matter to an energy and a density at saturation given by $E/A=-16$ MeV and $\rho_0 = 0.16$ fm$^{-3}$ and an incompressibility modulus  
$K_0 =230$ MeV. In the present case, each level is 4 time degenerated and the projection can be made separately on proton and neutron. 
Note that, calculations with Coulomb and spin-orbit can be made but we do not anticipate that the conclusion below will change. 
  
Again, for the sake of simplicity, we assume that the problem is first solved self-consistently 
without pairing, leading to a set of occupied single-particle states.  Our goal is then to vary at will 
the pairing gap, that will become for us a free parameter of the calculation. To do this, we consider 
the BCS approximation and neglect the effect of the pairing interaction on the single-particle energies. 
Then, for each couple of time-reversed degenerated states, we assign the $(u_n,v_n)$ components given by:
\begin{eqnarray}
u^2_n &=& \frac{1}{2} \left( 1+ \frac{(\varepsilon_n - \mu)}{E_n} \right), ~~
v^2_n= \frac{1}{2} \left( 1 -  \frac{(\varepsilon_n - \mu)}{E_n} \right)\nonumber
\end{eqnarray}
where $\mu$ is the chemical potential, while $E_n = \sqrt{(\varepsilon_i - \mu)^2 + \Delta^2}$ are the quasi-particle energies.
For each value of the gap, the chemical potential $\mu$ is adjusted such that the average number of 
particles equals the one we are interested 
in \footnote{Note that, in the absence of good reason to do so, 
we did not explore here the possibility to fix the average particle number to a value that would be different to the number 
of particles we are interested in. Therefore in this work the symmetry breaking 
state as always in average a number of particles that matches 
the total number of  particles $A$ we project onto.}, 
i.e. $\langle \hat A \rangle = A = 2 \sum_n v^2_n $.  Note that, to avoid difficulties with the continuum, only bound states contributes to the pairing correlations. 
In the following, we will show illustrations of counting statistics in a $^{40}$Ca nucleus. Its density is displayed in Fig. \ref{fig:densA40} for different pairing.  
In this figure and in the following, we will present densities, distances and gaps respectively in units of the saturation density 
$\rho_0$, average inter-particles distance $d$ and a quantity $\Delta \varepsilon$ related to the single-particle gap close to the Fermi energy.
The quantity $d$ is related to $\rho_0$ through $\rho_0 = 1/d^3$. We define the quantity $\Delta \varepsilon$ as    
\begin{eqnarray}
\Delta \varepsilon &=& \frac{1}{2} \left( \varepsilon_+ - \varepsilon_- \right)  \nonumber
\end{eqnarray}  
where $\varepsilon_-$ (resp. $\varepsilon_+$) corresponds to the energy of the level below (resp. above) the last  occupied level. For the $^{40}$Ca, this 
energy is given by $\Delta \varepsilon = 6.73$ MeV. In Figure \ref{fig:densA40}, we consider the two types of volume that will be considered in the present 
work. In one case we will consider the volume inside a sphere of radius $R$ while in the second case we will count particles outside the sphere. The former 
case, will be useful to study the counting statistics in the interior of the system while, in the former case, we will focus our attention on the surface properties.

\subsection{Counting statistic at the center of the system}

\begin{figure}[htbp]
\includegraphics[width=\linewidth]{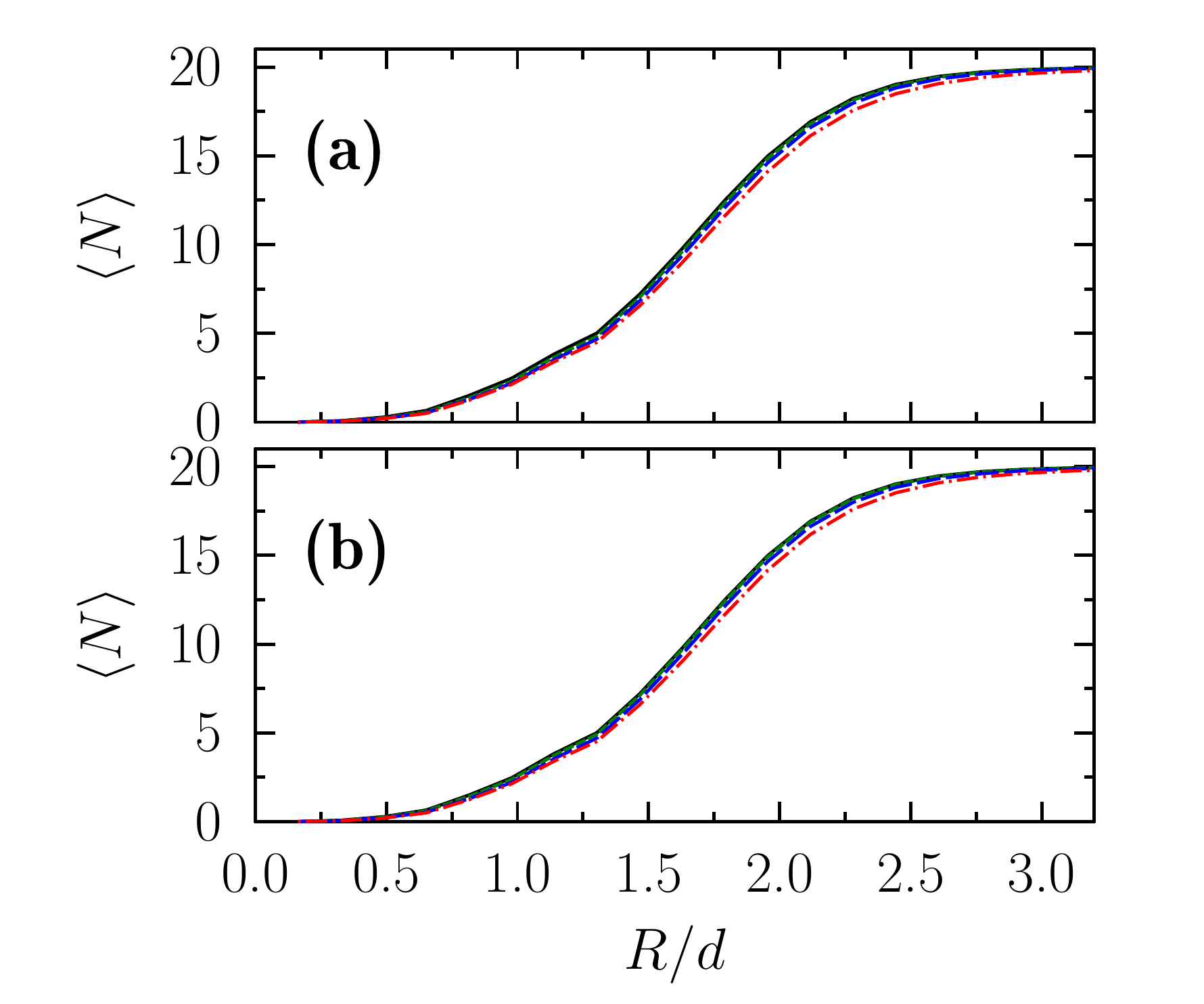}
\caption{Evolution of the mean number of particles in a sphere of radius $R$  ($|r| < R$)
 without (a) and with (b) projection on the total number of particles as a function of $R/d$. In both panels, the different lines 
 correspond to $\Delta/\Delta \varepsilon=0.01$ (black solid line),  $\Delta/\Delta \varepsilon=0.5$ (green short dashed line), $\Delta/\Delta \varepsilon=1.0$ 
(blue dashed line) and $\Delta/\Delta \varepsilon=2.0$ (red dot-dashed line).} 
\label{fig:mom0N40} 
\end{figure}

We first consider the case where the volume corresponds to a sphere of radius $R$ centered at the center of mass position of the system 
(blue shaded area on the left in Fig. \ref{fig:densA40}). In Fig. \ref{fig:proball}, we compare systematically the probability distribution $P(N)$ from 
very small volumes ($R/d=0.5$) to volumes comparable to the system size ($R/d=3.$).  
We first concentrate on the small volume limit. We see that the counting statistics in this case is almost independent on the fact that the 
projection on the total mass $A$ is made or not. In addition, it also depends  very weakly on the value of the gap itself. 
Such weak dependence can be 
directly attributed to the fact that in the limit of small volume, the counting statistics becomes a Poisson distribution that only depends on the mean value 
$\langle \hat N_\Omega \rangle$. The number of particles 
itself depends weakly on the presence of pairing or on the restoration of the total particle number (se Fig. \ref{fig:mom0N40}). It is however worth to mention 
that the weak dependence directly stems from the fact that the average number of particles $\langle \hat A \rangle$ used to fix the chemical potential $\mu$ is also 
equal to the number of particle on which the total projection is made. Using a different $\langle \hat A \rangle$ value, although not very natural, would lead to 
completely different evolution of $\langle \hat N_\Omega \rangle$ in the symmetry restored case compared 
to the symmetry breaking state.  

\begin{figure}[htbp]
\includegraphics[width=\linewidth]{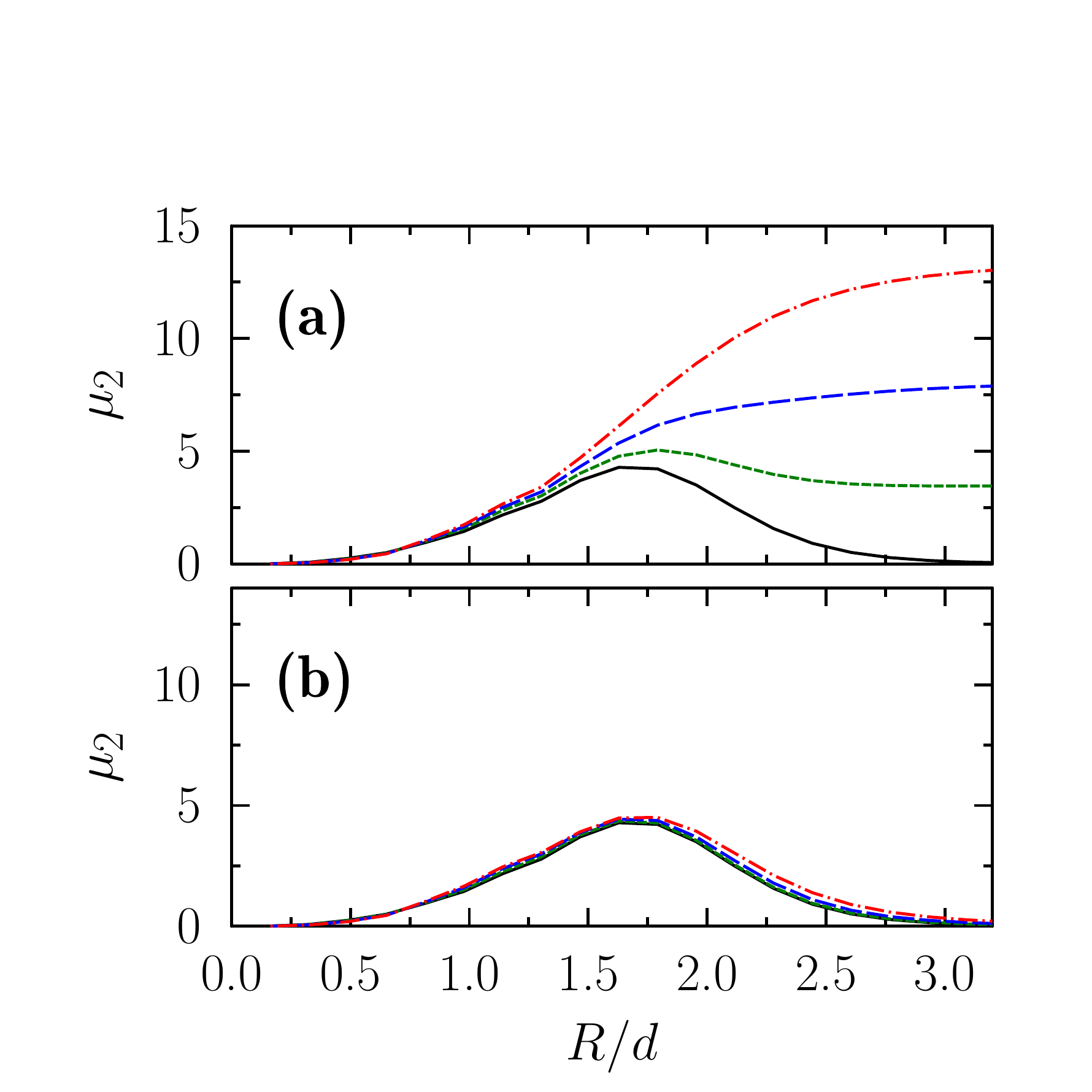}
\caption{Same as Fig. \ref{fig:mom0N40} for the second centered moment $\mu_2$.} 
\label{fig:mom2N40} 
\end{figure}

\begin{figure}[htbp]
\includegraphics[width=\linewidth]{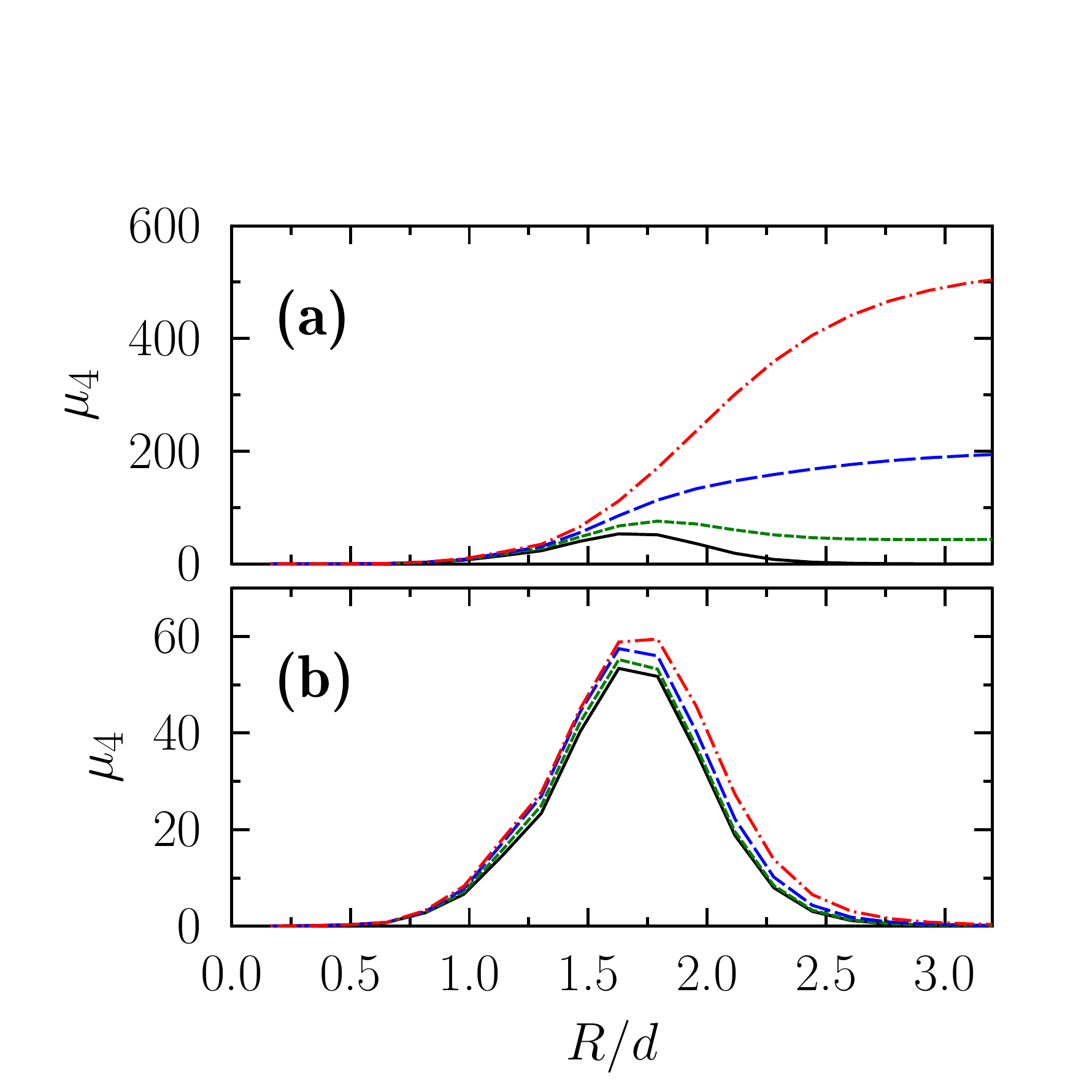}
\caption{Same as Fig. \ref{fig:mom0N40} for the fourth centered moment $\mu_4$.} 
\label{fig:mom4N40} 
\end{figure}

The situation is quite different for higher moments of the probability distribution. The second and fourth moments denoted by $\mu_2$ and $\mu_4$
are shown with and without restoration of the total particle number in Fig. \ref{fig:mom2N40} and \ref{fig:mom4N40} respectively. 
We clearly see in both figures that, in the absence of total particle number restoration, the centered
moments $\mu_2$ and $\mu_4$ have a significant dependence with the pairing field strength. More precisely, 
we observe that below $R/d=1$, these moments are almost not affected by the presence of pairing or by the projection 
while for larger volumes, the different moments increase significantly with the pairing gap. This dependence is a spurious effect that essentially    
stems from the use of a quasi-particle vacuum where the $U(1)$ symmetry is broken. Obviously, in a theory where the number of particle is preserved, 
all centered moments should go to zero when the volume becomes larger than the system size.  We see in panels (b) of Fig. \ref{fig:mom2N40} and 
\ref{fig:mom4N40} that this behavior is indeed observed when the symmetry is restored. While for small volumes projection onto total particle 
number does not affects the moments, for volumes with $R/d > 1$, the moments are strongly reduced after projection 
although a fossil dependence with the gap $\Delta$ persists.

The influence of the symmetry restoration on the counting statistics  is directly illustrated in Fig. \ref{fig:proball}  
and becomes more and more evident when either the gap or the volume increases. For instance, in the largest volume displayed 
in this figure ($R/d=3$), we clearly see the spurious contributions of the surrounding $A\pm2$, $A\pm4$, $\cdots$ nuclei.

Coming back to the small volume limit, it is interesting to mention that 
the fact that the counting statistics is almost unaffected by restoration of particle number can be seen 
as a further numerical proof of the intuition of Anderson that a small part of a system can be treated 
by a symmetry breaking state \cite{And66}. Such assumption was made for instance  in Ref. \cite{Bel07} without further justification to discuss 
the counting statistics. The occurrence of a Poisson distribution when the volume becomes very small is similar to the case 
of non-interacting particles (even treated as classical particles) when the volume becomes smaller to the volume 
occupied by a single particle \cite{Lan80} and directly reflects the independent quasi-particle nature assumption 
made for the trial state. As a consequence, the fluctuations of the local density are automatically divergent. This divergence is inherent to the 
use of quasi-particle vacuum. 

\subsubsection{Nature of the fluctuations} 
\begin{figure}[htbp]
\includegraphics[width=\linewidth]{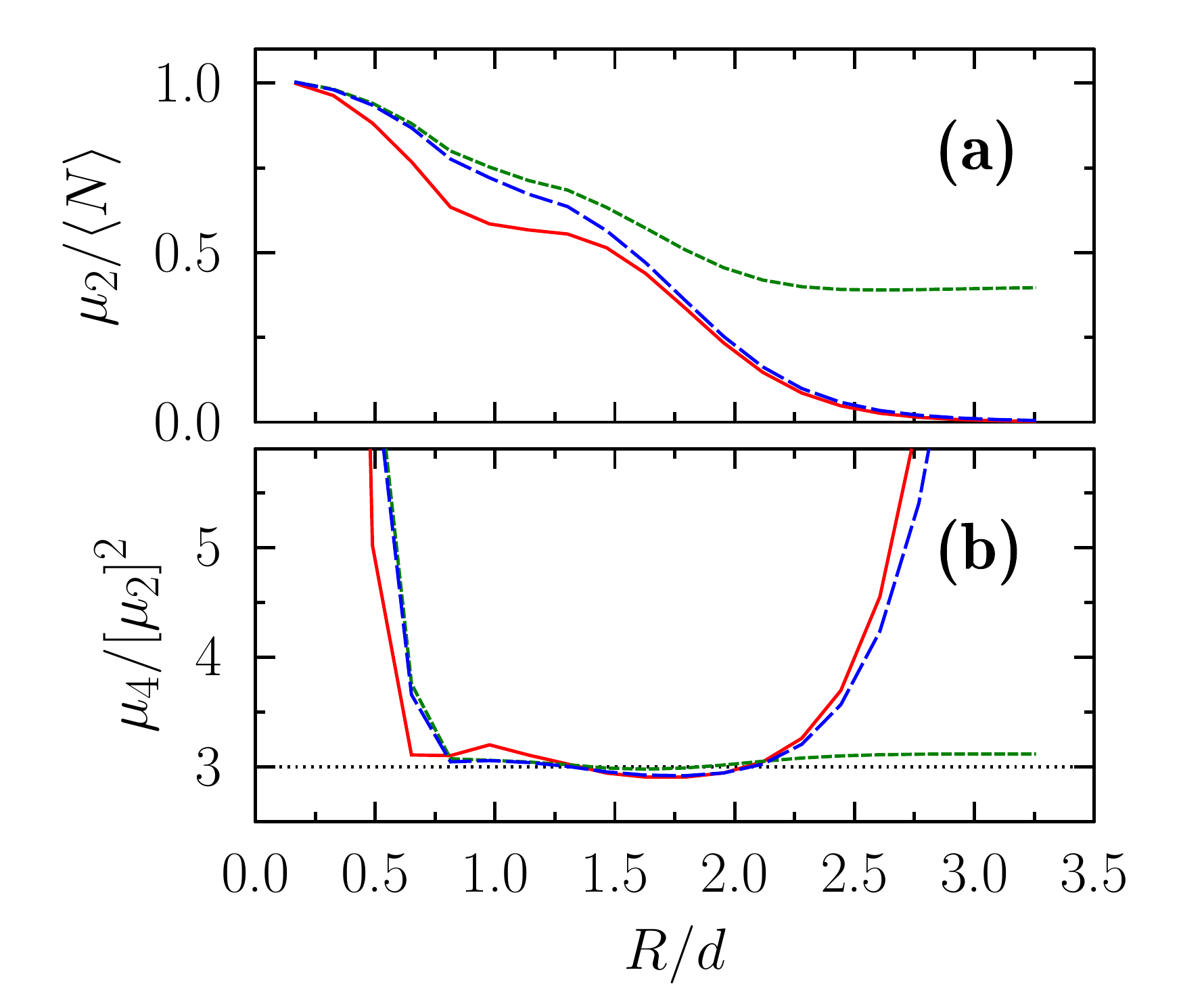}
\caption{Evolution of the quantities (a) $\mu_2/\langle N\rangle$ and (b)  $\mu_4/ \mu^2_2$ as a function of $R/d$. The case $\Delta/\Delta \varepsilon=0.01$ (red solid line), 
$\Delta/\Delta \varepsilon=1$ with single- (green short-dashed line) and double-projections (blue long-dashed line) are shown. The black dotted line in panel (b) corresponds to
the reference curve $\mu_4/ \mu^2_2=3.0$.} 
\label{fig:momN40} 
\end{figure}

We already know that the probability becomes a Poisson distribution when the volume is sufficiently small ($R/d<1$). In an infinite 
systems, when the volume increases, the number of particles participating to the probability increases, we expect that the probability 
distribution will become a Gaussian distribution due to the central limit theorem (see also discussion in section \ref{sec:diag}).
 
In large systems, the system will also reach the macroscopic regime where the fluctuations with respect to 
the mean becomes much smaller than the mean value itself. In a finite system, where the number of particles 
can be rather small, it is unclear if and when these two limits can be reached. For instance, the size of the system as well as the maximal number of particles 
at play in the fluctuations are strict boundary conditions for the volume and for the probabilities respectively. We then anticipate that this should affect 
the Gaussian nature of the probabilities when the volume becomes close to the system size.

To uncover the nature of the probability, we show in Fig. \ref{fig:momN40} the two ratios  $\mu_2/\langle N\rangle$ and  $\mu_4/ \mu^2_2$ for 
the case without pairing and for the case with pairing restoring or not the particle number symmetry. We do not show the third centered moments 
that is found very small when $\mu_4/\mu_2^2 \simeq 3 $. There are two limits that are interesting for us. The first one is the limit
$\mu_2 /\langle N \rangle \sim 1$ that corresponds to the Poisson probability. This limit is reached in all cases when the volume tends to zero and becomes 
an infinitesimally small volume ($R \ll d$). When the volume increases, this ratio decreases and we are always in the sub-Poissonian regime.   

The second limit is the case where $\mu_4 / \mu^2_2 \simeq 3$, that together with 
a vanishing third centered moment $\mu_3 \simeq 0$ is a strong indication of a Gaussian probability. As clearly seen in panel (b) of Fig. \ref{fig:momN40} for all 
cases, the ratio $\mu_4 / \mu^2_2 \simeq 3$ becomes very close to $3$ as soon as the radius approaches $R/d \simeq 1$. This is a direct signature that the 
probability becomes close to a Gaussian distribution as soon as very few particles contribute to the fluctuations.  In a symmetry conserving or symmetry 
restored approach, the Gaussian nature of the probability breaks down when the volume approaches the size of the system. Note that this is not the case 
when the particle number is not conserved where we see that the probabilities can still be approximated fairly well by a Gaussian at volume comparable 
or larger than the system size.

To further illustrate the Gaussian nature of the fluctuation, we also display in figure \ref{fig:proball}, 
the Gaussian probability $P_{\rm G}$ given by:
\begin{eqnarray}
P_G(N) &=& {\cal C}_\Omega e^{-(N-\lambda_\Omega)^2/(2\sigma^2_\Omega)} \label{eq:tdga} 
\end{eqnarray}
where $\lambda_\Omega$ and  $\sigma^2_\Omega$ identifies with the first and second moments obtained 
from the distribution we want to compare to. We assume that the Gaussian probability $P_G(N)$ is truncated in the sense that we do not allow for 
the possibility to have negative masses and also discretized since we only consider integer values for $N$. 
${\cal C}_\Omega$  is a normalization factor that insures $\sum_{N>0} P_G(N) =1$. This approximation is called hereafter 
the Truncated Discretized Gaussian Approximation  (TDGA).   The TDGA approximation obtained from the probabilities with and without 
projection onto the total mass is systematically shown with symbols in Fig.  \ref{fig:proball}.  We observe that, unless the 
volume becomes very small and/or comparable to the system size, the TDGA approximation becomes a very good approximation to the 
original probability. Note that we also tested the possibility to approximate the counting statistics obtained for 
various $\Omega$ either by a Poisson or by a binomial distribution, but such hypothesis 
do not reproduce the probabilities in general. Only, in the small volume limit, as expected, the Poisson probability provides the best approximation.  

Another approximation we explored was to neglect the off-diagonal matrix elements of the overlap matrix. In this case, as seen for instance in 
section \ref{sec:diag} and appendix \ref{app:double}, the Pfaffian calculation reduces to a simple product calculation. We have seen in general that 
this diagonal approximation can have significant deviations from the true probability.

\subsection{Counting statistic at the surface of the system}

We now consider a volume that will probe the counting statistics at the surface. More precisely, we consider a volume 
depicted by the yellow shaded area on the right in Fig. \ref{fig:densA40}  and that corresponds to the volume outside a sphere of radius $R$ ($|r|>R$). 
The counting statistics at the surface is of special interest for instance for nuclear reactions where a geometric picture 
of the reactions is often taken as it is the case for instance in the Glauber theory or more generally in the Abrasion-Ablation picture. 
This is also the assumption made quite often in fast peripheral collisions like in the knock-out reactions. In addition, we
expect that particles at the surface are mostly the ones with energy around the Fermi energy that are also usually the ones 
that are paired. For these reason, it is interesting to see if counting statistics behaves differently compared to the case with particles in the bulk where 
a fraction of the particles are unpaired.       
\begin{figure}[htbp]
\includegraphics[width=0.9\linewidth]{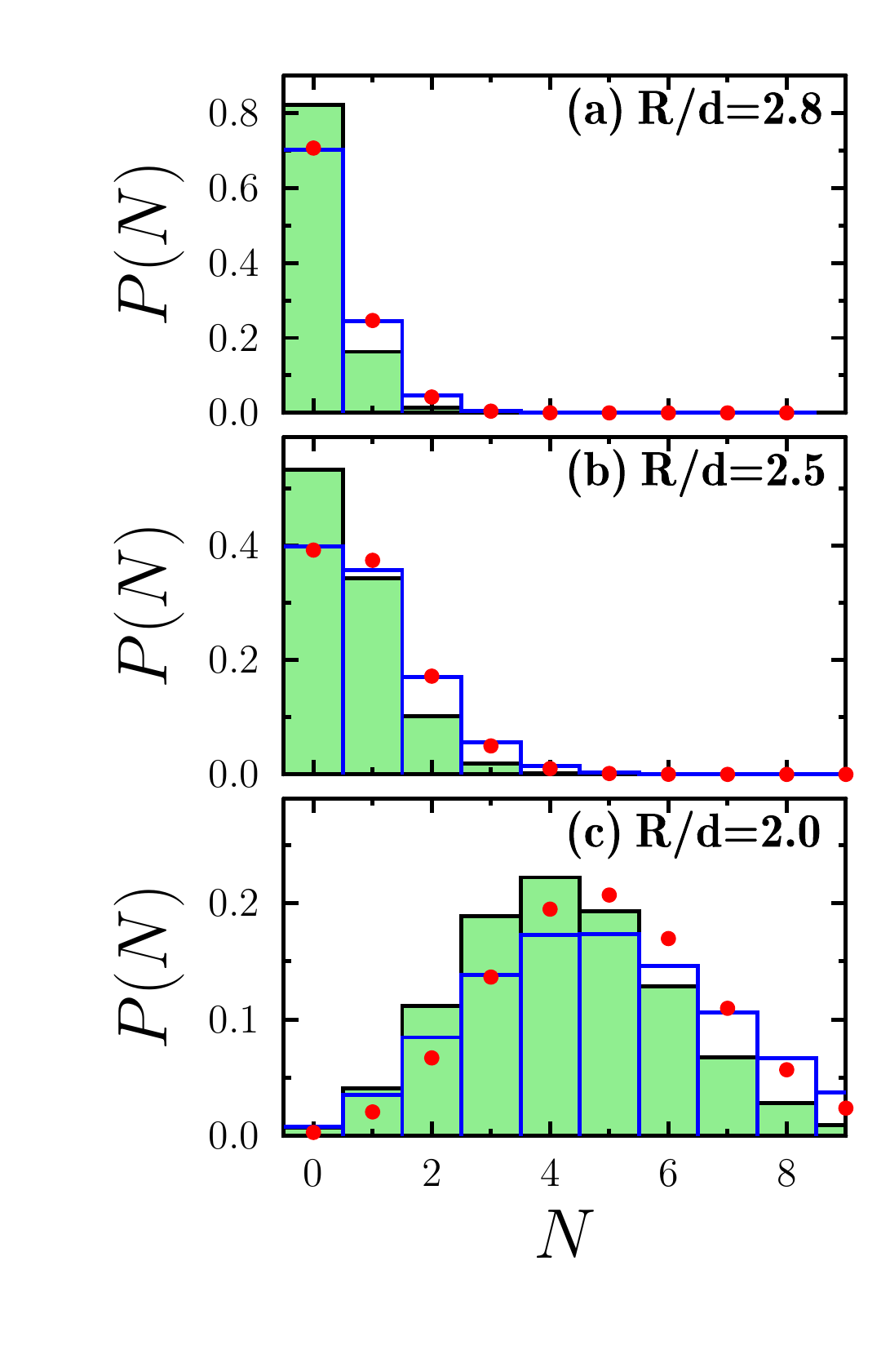}
\caption{Particle number probability distribution in the region $|r| > R$ for (a) $R/d=2.8$, (b) $R/d=2.5$ and (c) $R/d=2.0$. In each panel, 
the green shaded area corresponds to the no pairing case 
$\Delta/\Delta \varepsilon = 0.01$, while the blue solid line and red 
filled circles are obtained for  $\Delta/\Delta \varepsilon = 2.0$ respectively without and with total number of particles restoration. } 
\label{fig:probsurf} 
\end{figure}
\begin{figure}[htbp]
\includegraphics[width=\linewidth]{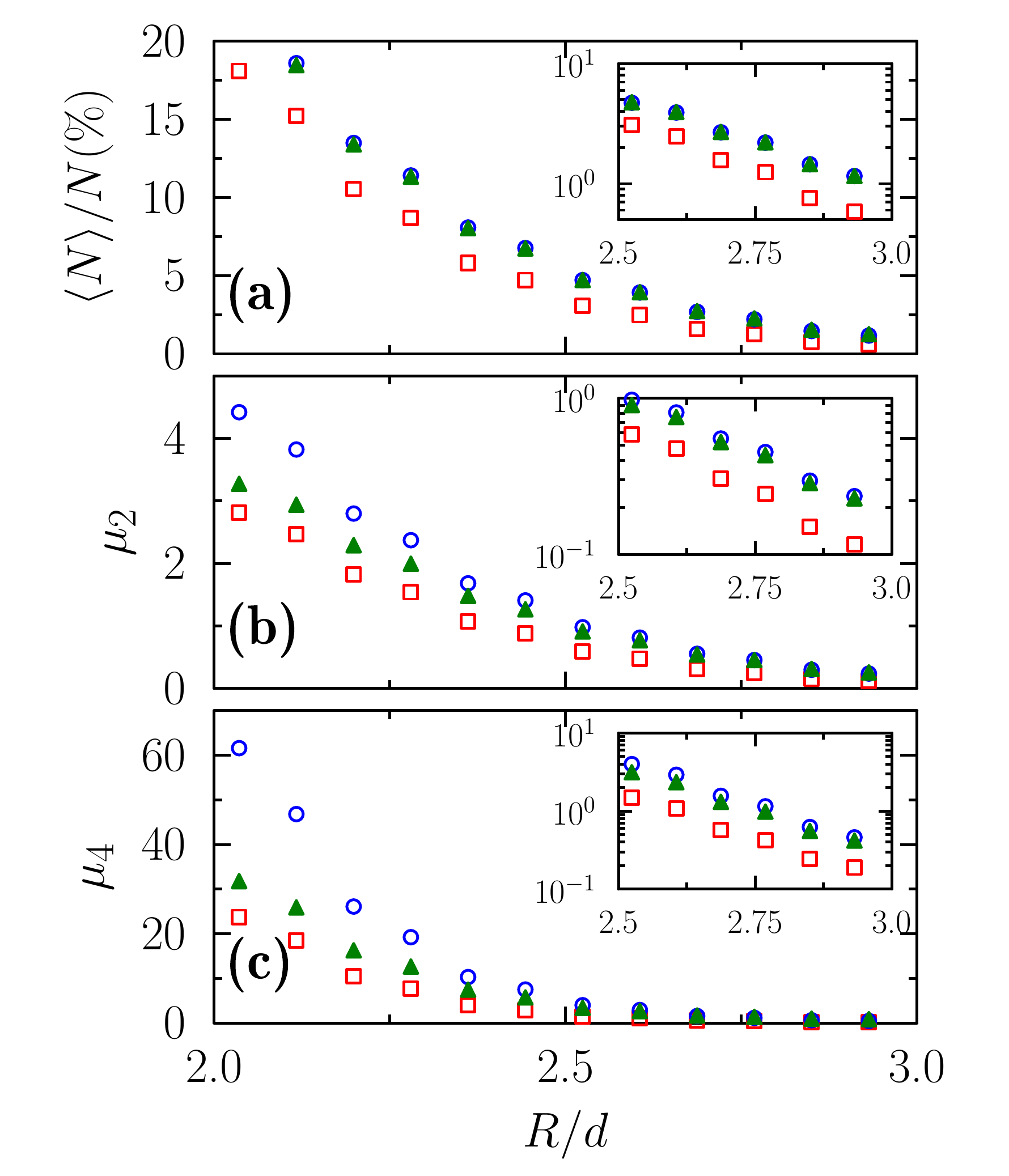}
\caption{ Evolution of (a) $\langle N\rangle/N$ (in \%), of the second (b) and fourth centered moment (c) as a function of $R/d$. 
In each panel, the case of $\Delta/\Delta \varepsilon=0.01$ (red open squares), 
$\Delta/\Delta \varepsilon=2.0$ with single- (blue open circles) or double- (green filled triangles) projection are presented. In each panel, the inset displays the 
large $R/d$ case in logscale at large values of $R$.
Here $R$ define the region such that  $|r| > R$. } 
\label{fig:mom4N40surf} 
\end{figure}

Illustrations of the counting statistics obtained by selecting different volumes that include more or less particles at the surface of the 
system are given in Fig. \ref{fig:probsurf} for the very small pairing case and for the case $\Delta/\Delta \varepsilon=2.0$ with and without restoration
of the $U(1)$ symmetry. 
To further characterize the probabilities, we also show in Fig. \ref{fig:mom4N40surf} the 
corresponding mean number of particles, second and fourth centered moments as a function of increasing volume (decreasing $R$).  

When the number of particles participating to the fluctuations in the volume $\Omega$ is small, as it is the case for $R/d=2.5$ and $R/d=2.8$, 
we observed essentially that (i) the change of the pairing strength modify slightly the probability (ii) restoring or not the total number of particles 
does not affect $P(N)$. This situation is similar to what we observed in panels (i) and (m) of Fig. \ref{fig:proball}. We therefore observed that this property
persists even at the surface when the density profile varies suddenly.

When $R$ decreases, i.e. when the volume increases and the number of particles participating to the fluctuations of $N$  in this volume increases, 
we observe that the width of the distribution after projections starts to deviate from the BCS case and approaches the case without superfluidity. 
This is again a direct effect of the total particle number projection that ultimately gives zero fluctuations when the volume $\Omega$ corresponds 
to the full volume.  Our conclusion is again that the restoration of the total number of particles is absolutely necessary when the number of particles 
involved in a process increases. In Fig. \ref{fig:mom4N40surf}, we see the onset of an effect of the $U(1)$ symmetry restoration  when $R/d<2.5$ 
which corresponds approximately to $5$ to $10$ $\%$ of the total number of particles $A=20$, i. e. $1$ to $2$ particles only. 
As a consequence, we anticipate that the quasi-particle approximation without restoration of $A$, that is for instance often used to 
describe grazing reactions, should be taken with some caution when more than one particle is involved in the process. This 
is for instance the case in the multi-nucleon transfer process studied with projection techniques that has been extensively 
discussed in recent years \cite{Sim10,Sca13,Sek14,Sek16,Has16,Mag17,Reg18,Reg19,Sca18}. 
This might even be more critical in fusion and/or fission process that reveal some global 
macroscopic transport properties of nuclei.  

As a final remark, we also mention that the analysis of the evolution of $\mu_2/\langle N\rangle$  and $\mu_4/\mu^2_2$ as a function of $R/d$
(not shown here) demonstrates that all probability distributions shown in Fig. \ref{fig:probsurf} tend to a Poisson distribution 
when the mean number of particles in the 
volume $\Omega$ becomes very small (here for $R/d \ge 2.8$), while it matches a Gaussian distribution when the number of particles increases 
(here $R/d\le 2.0$).     


\section{Conclusions}

In the present work, we have analyzed in details the evolution of the counting statistics in a finite volume inside a Fermi system taking the 
example of atomic nuclei. The nature of the probability distribution is studied as a function of the volume size and, as expected, we show 
that the probability distribution of particle number tends to a Poisson distribution when the volume becomes infinitesimally small. When the 
volume increases, the average number of particles participating to the fluctuations increases and as soon as more than one particle contribute
in average, the probability distribution becomes approximately Gaussian. These conclusions hold both when the volume probes the interior or the 
surface of the finite system and when the system is in a normal or superfluid phase.

In the superfluid regime, we use double-projection technique to restore the total number of particles and uncover possible effects
of the symmetry restoration on the counting statistics. We found that the restoration of particle number has a marginal effects 
when a small volume of the system is considered. This indirectly justify the use of BCS like or HFB states to describe physical process 
where one or maximum two particles are involved like it is the case sometimes in most peripheral reactions. However, when the 
volume increases, even in a regime where the macroscopic limit is far from being reached, the restoration of the particle number 
symmetry changes considerable the counting statistics. This implies that conclusion drawn using a symmetry breaking state 
on a process where several particles are simultaneously involved  should be taken with care unless the total number of particle is properly 
restored. Examples of such processes are deep inelastic collisions or fission. 

As a final remark, we would like to mention that the counting statistics obtained here stems from a nuclear  density functional theory
where the system is described by an independent quasi-particle vacuum. In particular, this counting statistics might differ from 
the one one would obtain from an ab-initio method where a complex many-body wave-function is used to solve the static problem starting from 
the bare Hamiltonian. Extracting the counting statistics in such a case might be very interesting. In particular, we anticipate differences 
with the present work due to the possible strong short-range correlations \cite{Cru18,Wei19a} that are linked to the Tan's contact parameter 
\cite{Tan08a,Tan08b,Tan08c,Bra12} in some cases.
These short-range physics is however out of the scope of a DFT approach.  

\begin{acknowledgments}
We thank D. Regnier for useful discussion and for his remarks on the manuscript. 
S.A. gratefully acknowledges the IPN-Orsay for  the warm hospitality extended to him during his visits. 
This work is supported in part by the US DOE Grant No. DE-SC0015513. This project has received 
funding from the European Union's Horizon 2020 research and innovation programme under Grant Agreement No. 654002.
\end{acknowledgments}

\appendix

\section{Simple cases illustrating the effect of total particle number projection on the counting statistics}
\label{app:double}

In the present section we consider simple situations that illustrate the effect of restoring the total particle number on the counting 
statistics. These examples although rather schematic can be of interest in nuclear physics in specific situations, like for instance when a single pair, or 
a single $j$-shell contributes to a physical process. 

\subsection{The case of a diagonal overlap}

Following section \ref{sec:diag}, we assume that the overlap matrix is diagonal, then again the matrix  ${\cal M}$ becomes 
a $4\times4$ bloc matrix. Accounting for the additional angle $\theta$ associated to the total particle number conservation, 
the generating function (\ref{eq:gen2}) is given by:
\begin{eqnarray}
F(\varphi,\theta) &=&  \prod_{n>0} \{ u^2_n  + v^2_ne^{2i\theta}[1+ p_n (e^{i\varphi} - 1)]^2\}. \label{eq:gentp}
\end{eqnarray}
From this expression, one can define for instance the generating function $\widetilde F(\varphi)$ associated to the 
probability to have $N$ particles in a volume $\Omega$ once the projection on the total particle number is performed. This generating
function is defined as:
\begin{eqnarray}
\widetilde F(\varphi) &=& \int \frac{d\theta}{2\pi} e^{-iA \theta} F(\varphi,\theta). \label{eq:redgen}
\end{eqnarray} 
Starting from Eq. (\ref{eq:gentp}), an explicit form of this generating function can be obtained by developing the product and by 
selecting the terms proportional to $e^{iA\theta}$. This will give a rather complicated expression that amount to count the number of ways 
to select $K=A/2$ different pairs into the total number $N_p$ for partially occupied pairs.   

\subsection{The case of a fully degenerated $j$-shell}

Let us assume that all particles belongs to a shell $(j,m)$ where all single-particle states are degenerated.
This is a simple situation were both the BCS or projected BCS states are known explicitly \cite{Bri05}. In this case, 
we have for all states $u_n=u$, $v_n=v$ and $p_n=p$ and the generating function both with and without projection 
onto good particle number can be worked out analytically. 

\subsubsection{Probabilities without restoration of the $U(1)$ symmetry breaking}

If the projection is applied only to the small volume, the partition function becomes:
\begin{eqnarray}
F(\varphi) 
&=&  \left[ u^2 + v^2 \left( q+ p  e^{i\varphi} \right)^2  \right]^{N_p}.\nonumber 
\end{eqnarray}

Let us introduce the notation  ${\cal P}= v^2$ and ${\cal Q}=u^2$. 
In this simple example ${\cal P}$ (resp. ${\cal Q}$) can be interpreted as 
the probability to occupy a pair. Then, one might denote by ${\bf P}_K$ the probability 
to occupy $K$ pairs among $N_p$. This probability is given by:
\begin{eqnarray}
{\bf P}_K = C^K_{N_p}{\cal Q}^{N_p - K}  {\cal P}^{K} . \nonumber
\end{eqnarray}
Using this notation, we can rewrite the generating function as:
\begin{eqnarray}
F(\varphi) &=& \sum_{K=0}^{N_p} {\bf P}_K \left( q+ p  e^{i\varphi} \right)^{2K}. \nonumber
\end{eqnarray} 
In each term, we recognize the generating function of the binomial distribution. Introducing 
the notation:
\begin{eqnarray}
B_{2K} (m) = C^{m}_{2K} p^{m} q^{2K-m}   ,
\end{eqnarray}     
we finally deduce that the probability to have $m$ particles in the small volume can be written as:
\begin{eqnarray}
P(m) =  \sum_{K=0}^{N_p} {\bf P}_K B_{2K}(m) . \nonumber
\end{eqnarray}  
This probability can be interpreted as the convolution between the probability to have $K$ occupied pairs 
with the probability within the $2K$ particles forming the pairs that $m$ particles are located 
in the volume $\Omega$.    
  
\subsubsection{Probabilities with restoration of the $U(1)$ symmetry breaking}
 
 When the restoration of the total particle number $A$ is accounted for, the partition function becomes:
 \begin{eqnarray}
F(\varphi, \theta) &=& \left[ u^2  + v^2 e^{2i\theta} \left( q  + p e^{i\varphi}\right)^2 \right]^{N_p} . \nonumber 
\end{eqnarray}
A straightforward calculation shows that the generating function (\ref{eq:redgen}) can be recast as (using the notation $A=2K$):
\begin{eqnarray}
\widetilde F(\varphi) &=& {\bf P}_K  \left( q  + p e^{i\varphi}\right)^{2 K} .\nonumber
\end{eqnarray}  
Using the normalized probabilities as given by Eq. (\ref{eq:probnorm}) will remove the completely the ${\bf P}_K$ coefficient and we finally obtained
that the probability to have $m$ particles identifies with the binomial law:
\begin{eqnarray}
P (m) &=& B_{2K}(m) .  \nonumber  
\end{eqnarray}
This example illustrates both the spurious contribution that might stems from the $U(1)$ symmetry breaking and 
the effect of restoring the total particle number.

\end{document}